\documentclass[prb,showpacs,aps,twocolumn]{revtex4}

\usepackage{amsmath, amssymb}
\usepackage{graphicx}
\usepackage{graphics}
\newcommand{\tr}{\mathop{\mathrm{Tr}}\nolimits}
\begin{document}
\title{Phonon-mediated decoherence in triple quantum dot interferometers}

\author{Fernando Dom\'inguez}
\author{Sigmund Kohler}
\author{Gloria Platero}
\affiliation{Instituto de Ciencia de Materiales, CSIC,
	Cantoblanco, E-28049 Madrid, Spain}

\date{\today}

\pacs{
      73.23.-b,	
      05.60.Gg	
      74.50.+r,  
}

\begin{abstract}

We investigate decoherence in a triple quantum dot in ring
configuration in which one dot is coupled to a damped phonon mode,
while the other two dots are connected to source and drain,
respectively.  In the absence of decoherence, single electron
transport may get blocked by an electron falling into a superposition
decoupled from the drain and known as dark state.  Phonon-mediated
decoherence affects this superposition and leads to a finite current.
We study the current and its shot noise numerically within a master
equation approach for the electrons and the dissipative phonon mode.
A polaron transformation allows us to obtain a reduced equation for
only the dot electrons which provides analytical results in agreement
with numerical ones.

\end{abstract}

\maketitle

\section{Introduction}

\begin{figure}[b]
\begin{center}
\includegraphics{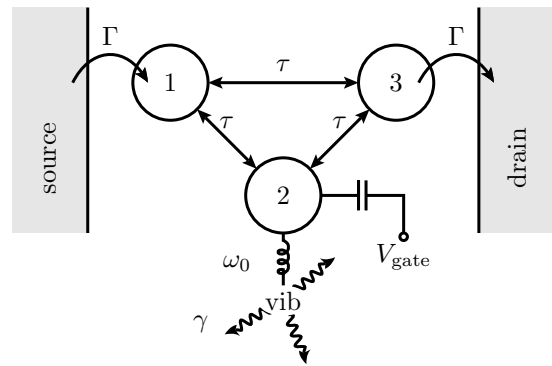}
\end{center}
\caption {\label{fig:setup}
Triple quantum dot in ring configuration with mutual tunnel couplings
$\tau$.  Dots~1 and 3 possess onsite energies $\epsilon_{1,3} = 0$ and
are tunnel coupled to the source and the drain, respectively.  Dot~2
interacts with a damped vibrational mode with frequency $\omega_0$,
while its onsite energy can be tuned by a gate voltage such that
$\epsilon_2 = V_{\rm gate}/e$. Dot~2 has a vibrational degree of
freedom, while dots~1 and~3 are rigidly attached to the contacts.}
\end{figure}

Coherently coupled quantum dots allow the experimental investigation
of electron transport through delocalized orbitals and the associated
coherent superpositions.  The latter are visible in the charging
diagram of double or triple quantum dots as broadened lines between
regions in which an electron is localized in the one or the other dot.
The consequence for the current-voltage characteristics is that
Coulomb steps discern into multiple steps, each corresponding to an
orbital that enters the voltage window.\cite{Gaudreau2006a,
Schroer2007a, Onac2006a, Taubert2008a} When coupled quantum dots are
arranged in a ring configuration as sketched in Fig.~\ref{fig:setup},
electrons can proceed in two ways from the source to the
drain.\cite{Gustavsson2006a, Rogge2008a}  Then interference effects
emerge, provided that the tunneling is coherent.  For cetain phases of
the tunnel matrix element, a superposition decoupled from the drain is
formed such that an electron may become trapped in the
interferometer.\cite{Michaelis2006a, Emary2007b, Busl2010a,Busl2010b}
Owing to Coulomb repulsion, these so-called dark states block the
electron transport.  Detuning the energy of one of the dots forming
the superposition resolves this blockade, but leads to temporal
trapping by off-resonant tunneling to and from  the detuned dot.  This
leads to avalanche-like transport with super-Poissonian
noise.\cite{Emary2007b, Dominguez2010a}

The natural enemy of interference is decoherence, i.e., the loss of
the quantum mechanical phase.  The common scenario for this process is
that the considered system interacts with environmental degrees of
freedom and, thus, becomes entangled with them.  Then tracing out the
environment diminishes interference and the system tends to behave
classically.  A frequently employed model for describing decoherence
is the linear coupling of a central system to a bath of harmonic
oscillators representing, e.g., phonons or photons.
\cite{Magalinskii1959a, Caldeira1983a, Leggett1987a, Hanggi1990a}
Owing to the linearity of both the bath and its coupling to the
system, the former can be eliminated\cite{Feynman1963a} yielding a
master equation or a path integral description of the now dissipative
central system.  If decoherence stems from the coupling to fermionic
baths such as nuclear spins or defects, a spin bath model is more
appropriate.\cite{Stamp1988a, Shao1998a, Prokofev2000a}  Electron spin
decoherence is can be induced by hyperfine interaction of an electron
placed in a single \cite{Yao2007a} or double quantum dot, where
decoherence affects spin blockade regime.\cite{Dominguez2009a}

A slightly different scenario is the so-called ``third-party decoherence''
\cite{Stamp2006a} in which a quantum system couples via a further
small quantum system to a bath consisting of many degrees of
freedom.  A particular case is the coupling of the quantum system via
a harmonic oscillator to a bath of harmonic oscillators.  This
system-oscillator-bath model is equivalent to a system-bath model with
a spectral density peaked at the oscillator
frequency,\cite{Wilhelm2004a, Thorwart2004a, Goorden2004a}  unless
nonlinearities of the oscillator are taken into
account.\cite{Vierheilig2009a}

Here we investigate how destructive interference in a triple quantum
dot interferometer is modified by the coupling to a dissipative
harmonic oscillator.  We focus on the regime of weak dot-lead
tunneling in which a master equation description is appropriate.
Nevertheless, the electron dynamics may exhibit non-Markovian effects
stemming from the coupling to the oscillator.  Therefore, it is
technically advantageous not to eliminate the oscillator but to treat
ii as part of the central system.

Our paper is organized as follows. In Sec.~\ref{sec:model} we
introduce the phonon-system-lead Hamiltonian and derive a quantum
master equation with which we investigate in Sec.~\ref{sec:numerics}
the impact of decoherence on the current and its noise.
Section~\ref{sec:limits} is devoted to an effective master equation
for only the dot electrons based on a polaron transformation.  Some
technical details of the derivation of the effective master equation
and the computation of the oscillator correlation function are
deferred to the appendix.


\section{Triple quantum dot in ring configuration}
\label{sec:model}

We consider three quantum dots in the ring configuration sketched in
Fig.~\ref{fig:setup}.  The electronic part consists of three quantum
dots that are mutually tunnel coupled.  Since we will focus on
decoherence effects stemming from the interaction with a phonon mode,
we neglect the spin degree of freedom.  Moreover, we restrict
ourselves to the limit of strong inter-dot and intra-dot Coulomb
repulsion such that only the states with zero or one excess electron
on the ring are relevant.  Thus, the only relevant states are the
empty state $|0\rangle$ and the one-electron states $|i\rangle =
c_i^\dagger|0\rangle$, where $i=1,2,3$ refers to the dot on which the
electron resides and $c_i^\dagger$ is the associated electron creation
operator.  Then the electronic part of the Hamiltonian reads
\begin{align}
H_\text{TQD} = \sum_{i=1}^3 \epsilon_i n_i 
+\tau \sum_{i> j} (c_i^\dagger c_j + \text{h.c.}) \,,
\label{eq:TQD}
\end{align}
where $\tau$ is the tunnel matrix element between dots $i$ and $j$, 
and $n_i=c_i^\dagger c_i$ the occupation number of the dot~$i$.
We consider the situation in which dots 1 and 3 are degenerate and
possess onsite energies $\epsilon_1=\epsilon_3=0$.  By contrast dot~2,
placed in one path of the interferometer, shall be tunable
by a gate voltage such that $\epsilon_2 = eV_\text{gate}$.
In order to include the Aharonov-Bohm phase produced 
by a flux $\Phi$ through the ring,\cite{Aharonov1959a} we multiply the
operators for clockwise tunneling by $e^{i\phi}$, while
counter-clockwise tunnel matrix elements acquire the factor
$e^{-i\phi}$, where $\phi = 2\pi\Phi/\Phi_0$ with the flux quantum
$\Phi_0 = h/2e$.

Dots 1 and 3 are tunnel coupled to metallic leads which is described by
the Hamiltonians
\begin{align}
H_\text{leads}
=& \sum_{\ell,k} \epsilon_{\ell k} c_{\ell k}^{\dagger}c_{\ell k},
\label{eq:ereservoir}
\\
H_\text{dot-leads} =& \sum_k (V_{L k} c_{L k}^\dagger c_1
                           +V_{R k} c_{ R k}^\dagger c_3 +\text{h.c.}),
\label{eq:saltos}
\end{align}
where $c_{\ell k}^\dagger$ and $c_{\ell k}$, $\ell = L,R$, create and
annihilate an electron in left and in the right lead, respectively.
The tunnel matrix elements $V_{\ell k}$ enter only via their spectral
density $\Gamma_\ell = 2\pi \sum_k |V_{\ell k}|^2
\delta(\epsilon-\epsilon_{\ell k})$ which we assume to be independent
of the energy $\epsilon$.  Then $\Gamma_\ell$ is the tunnel rate
between lead $\ell$ and the respective dot.

\subsection{Electron-phonon interaction}

An electron on dot~2 interacts linearly with a localized
 phonon mode according to \cite{Brandes2003a} 
\begin{align}
&H_\text{ph} =\hbar\omega_0 a^\dagger a ,\\
&V_\text{e-ph}=\lambda c_2^\dagger c_2 (a^{\dagger}+a) ,
\label{eq:intep}
\end{align}
which can be interpreted as a dynamical energy shift.  In turn, an
electron on dot~2 entails a force on the oscillator, such that the
latter acquires information about the path that an electron takes on
its way from source to drain.  Such ``which way information''
influences interference properties.  Notice that we treat the coupling
energy $\lambda$ as parameter despite the fact that it can be
determined from microscopic considerations.\cite{Brandes2003a}

Dissipation of the localized phonon mode $a$ stems from the interaction
with a bosonic environment such as substrate phonons.  The
environment and its coupling to mode $a$ are described by the
system-bath Hamiltonian
\begin{align}
H_\text{env} =& \sum_\nu\hbar \omega_\nu a_\nu^{\dagger}a_\nu,
\label{eq:sm}
\\
H_D =& (a^\dagger+a) \sum_\nu \lambda_\nu (a_\nu^{\dagger}+ a_\nu) ,
\label{eq:damping}
\end{align}
where $a_\nu$ and $a_\nu^{\dagger}$ are the creation and annihilation 
operators of the bath modes, while $\lambda_\nu$ are the coupling
constants.  The influence of the environment is fully determined by its
spectral density $I(\omega) = \pi \sum_\nu |\lambda_\nu|^2
\delta(\omega-\omega_\nu)$, which we assume to be Ohmic,
i.e., $I(\omega) = \gamma\omega$, where $\gamma$ denotes the effective
damping rate.

\subsection{Quantum master equation}

In order to derive a master equation for the dissipative dynamics of
the triple quantum dot and the localized mode, we start from the
Liouville-von Neumann equation for the full density operator,
$i\hbar\dot R =[H_{\text{tot}},R]$, 
where $H_{\text{tot}}$ is the sum of all the Hamiltonians appearing above.
  Using standard techniques,\cite{Blum1996a} we
obtain for the reduced density operator the equation of motion
\begin{align}
\label{ME-rho}
\dot \rho
={} & -\frac{i}{\hbar}[ H_0, \rho]
\nonumber
\\ &-\frac{1}{\hbar^2}
   \tr_\text{leads+bath} \int_0^\infty d t
  [H_V,[\tilde H_V(-t), R]]
\\
\equiv{} & \mathcal{L}\rho ,
\end{align}
which can be evaluated under the factorization assumption $R \approx
\rho_{\text{leads},0} \otimes \rho_{\text{bath},0} \otimes\rho$.
We have defined $H_0 = H_\text{TQD} +H_\text{ph} +V_\text{e-ph}$.
The tilde denotes the interaction picture with respect to the
Hamiltonian $H_0 + H_\text{\text{leads}}+H_\text{\text{bath}}
$, i.e., $\tilde X(t) = U_0^\dagger(t) X U_0(t)$.
The coupling of the central system to the leads and the heat bath
has been subsumed in the interaction Hamiltonian $H_V =
H_\text{dot-leads} + H_D$.

We insert $H_\text{dot-leads}$ and $H_{D}$ and evaluate the trace of
the electron and phonon reservoirs to obtain the Liouvillian
\cite{Armour2002a, Gurvitz1996a}
\begin{equation}
\begin{split}
\mathcal{L}\rho
=& -\frac{i}{\hbar}[ H_0, \rho]
   -\frac{\Gamma_L}{\hbar} (2 c_1\rho c_1^{\dagger}-c^{\dagger}_1
    c_1\rho-\rho c^{\dagger}_1c_1)
\\
& -\frac{\Gamma_R}{\hbar} (2 c_3\rho c_3^{\dagger}-c^{\dagger}_3
   c_3\rho-\rho c^{\dagger}_3c_3 )
\\
& +\frac{\gamma}{2}(\bar n+1)(2a\rho a^{\dagger}-a^{\dagger}a \rho
  -\rho a^{\dagger}a)
\\
& + \frac{\gamma}{2}\bar n (2 a^{\dagger}\rho a-aa^{\dagger}\rho
  -\rho a a^{\dagger}) ,
\label{eq:TrME}
\end{split}
\end{equation}
where $\bar n = [\exp(\hbar\omega_0/k_BT)-1]^{-1}$ is the thermal
occupation number of the localized mode at temperature $T$.
Restricting ourselves to the limit in which all dot states lie within
the voltage window, we have replaced the Fermi function of the left
lead by 1 and that of the right lead by 0.  Only in this limit, the
dot-lead tunnel terms proportional to $\Gamma_{L,R}$ assume this
simple form.  Moreover, we consider the oscillator dissipation within
rotating-wave approximation.\cite{Gardiner2004a}

In order to obtain a current operator in the reduced Hilbert space, we
start from the definition of the current as the change of the charge
in the right lead $e N_R$.  The according current operator $\mathcal{J} =
(ie/\hbar)[H_{\text{tot}},N_R]$ still depends on lead operators.  These are
eliminated within the same approximations that yield the master
equation~\eqref{eq:TrME}.  The result can be separated into two
contributions, $\mathcal{J}^+$ and $\mathcal{J}^-$, which describe
electron tunneling from the triple quantum dot to the right lead and
back, respectively.\cite{on_currentoperator}\nocite{Kaiser2007a} In
the present case of unidirectional transport, $\mathcal{J}^-=0$, while
\begin{equation}
\mathcal{J}^+ = \frac{e\Gamma_3}{\hbar} c_3^{\dagger}\rho c_3 .
\end{equation}
Then the stationary current expectation value reads
\begin{equation}
\label{current}
I = \tr\mathcal{J}^+\rho_\infty ,
\end{equation}
where $\rho_\infty$ denotes the stationary solution of the master
equation \eqref{eq:TrME}.

Further information about the transport process is provided by the
zero-frequency noise $S$ which is essentially the rate at which the
charge variance in one lead changes, i.e., $S =
\lim_{t\to\infty}\langle\Delta Q_R^2\rangle/t$.  It can be computed in the
same way as the stationary current but with $N_R$ replaced by $N_R^2$.
For unidirectional transport, one obtains\cite{Novotny2004a}
\begin{equation}
S = e\tr\mathcal{J}^+\rho_\infty - 2e\tr\mathcal{J}^+
{\hat{\mathcal{L}}}^{-1}\mathcal{J}^+\rho_\infty ,
\end{equation}
where ${\hat{\mathcal{L}}}^{-1}$ is the pseudo-inverse of $\mathcal{L}$, whose
action on $\mathcal{L}\rho_\infty \equiv X$ is computed by solving
$\mathcal{L}X = \mathcal{J}^+\rho_\infty$ under the condition $\tr
X=0$.  Below we will always discuss the noise strength in relation to
the current.  This motivates the definition of the Fano factor $F =
S/eI$, which assumes the value $F=1$ for a Poisson process.

For a numerical solution, we will have to truncate the Hilbert space
of the localized phonon mode at some maximal phonon number $N$.
Unless explicitly stated otherwise, truncation at $N=20$ ensured
numerical convergence.

\section{Transport properties: numerical results}
\label{sec:numerics}

In order to outline the behavior of the triple dot under the influence
of the dissipative phonon, we investigate numerically two situations.
In the first one, all dots are in resonance, such that a dark state
blocks transport.  The second one is that of a strongly detuned dot~2,
in which the blocking becomes imperfect.  We also consider a magnetic
flux through the triple quantum dot for ascertaining interference.

\subsection{All dots in resonance}

\begin{figure}[tb]
\begin{center}
\includegraphics{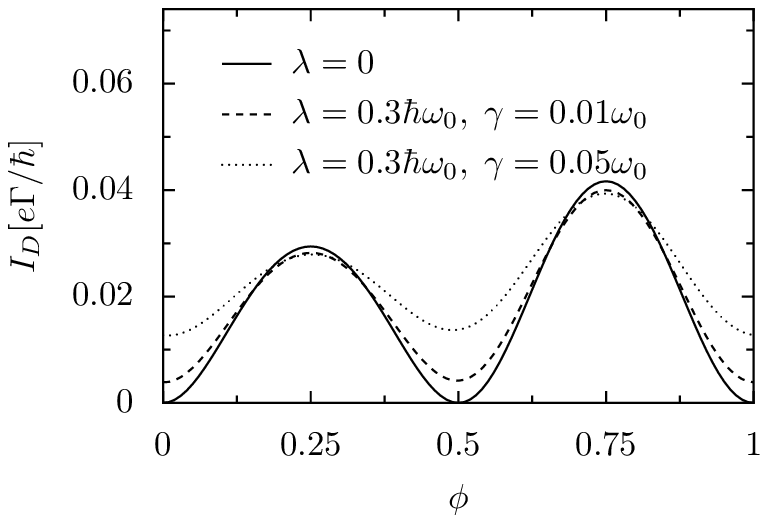}
\end{center}
\caption {\label{Fig.oscilaciones}
Current at zero detuning, $\epsilon_2=0$, as a function of the scaled
magnetic flux $\phi$ for two values of the phonon damping strengths
$\gamma$ compared to the current in the absence of the phonon
($\lambda=0$).  The dot-lead tunneling rate is $\Gamma =0.1\omega_0$.
In the non-interacting case, the current drops to zero at semi-integer
values of the quantum flux. The Aharonov-Bohm amplitude is reduced by
the phonon-mediated decoherence of the dark state.}
\end{figure}%
For a small gate voltage such that $|\epsilon_2| \ll \tau$ all three
dots are near resonance, and therefore interference is important.  For
the present configuration in which all three inter-dot tunnel couplings
are equal, it has been shown that for $\epsilon_2=0$ an electron is
trapped in the superposition\cite{Michaelis2006a, Emary2007b}
\begin{equation}
|\Psi_{\text{dark}}\rangle=\frac{1}{\sqrt{2}}(|1\rangle-|2\rangle).
\label{eq:darkwf}
\end{equation}
Obviously, it is orthogonal to state $|3\rangle$ and, thus, is
decoupled from the drain.  This implies that once an electron
populates state \eqref{eq:darkwf}, it cannot leave the triple dot.
Since Coulomb repulsion inhibits further electrons from entering the
dots, the current vanishes.  At zero flux, $\phi=0$, the two paths
$|1\rangle \rightarrow |3\rangle$ and $|1\rangle \rightarrow
|2\rangle\rightarrow |3\rangle$ interfere destructively at the
drain.\cite{Michaelis2006a, Emary2007b}  If $\phi$ is changed, a
finite current flows, unless $\phi$ assumes a semi-integer
value,\cite{Emary2007b, Busl2010a} as is visible from the
Aharonov-Bohm oscillations depicted in Fig.~\ref{Fig.oscilaciones}.
Figure~\ref{Fig.oscilaciones} also shows that when coupling dot~2 to
the oscillator, Aharonov-Bohm oscillations fade out with increasing
dissipation strength $\gamma$, which is a signature for the influence
of decoherence.  Moreover, it can be seen that this fading can be read
off faithfully at $\phi=0$ and, thus, henceforth we restrict ourselves
to this value.

\begin{figure}[tb]
\begin{center}
\includegraphics{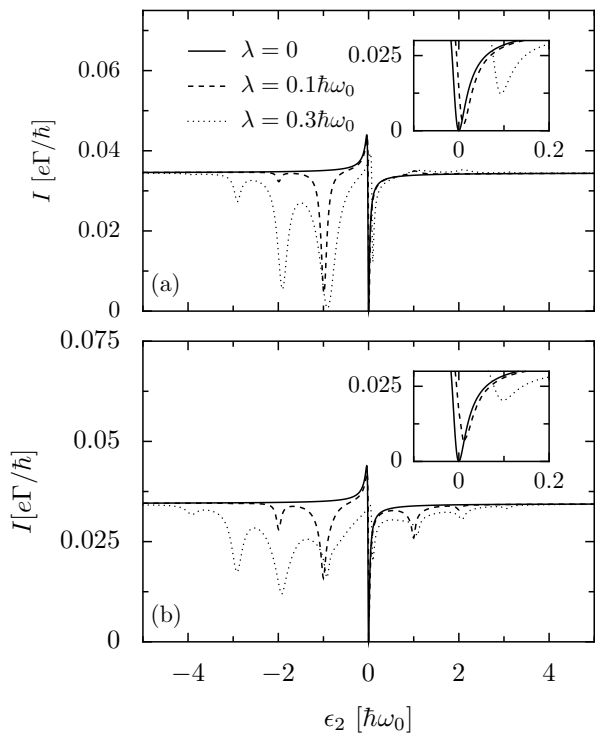}
\end{center}
\caption {\label{Fig.current} Current as a function of the detuning
$\epsilon_2$ for various values of the electron-phonon coupling
$\lambda$. The inter-dot tunneling and the dot-lead tunneling rate are
$\tau=0.01\hbar\omega_0$ and $\Gamma =0.1\omega_0$, respectively,
while the dissipation strength is $\gamma=0.05\omega_0$.  The
temperatures are (a) $T=0 $ and (b) $T=1.5\hbar\omega_0/k_B$.  Insets:
enlargement of the region near $\epsilon_2=0$ demonstrating a shift of
the current minimum with increasing electron-phonon coupling.}
\end{figure}%
The insets of Fig.~\ref{Fig.current} show the current as a function of
the detuning for various electron-phonon coupling strengths $\lambda$
and two different temperatures for small detuning.  An interesting
observation is that with increasing electron phonon coupling (see
insets of Fig.~\ref{Fig.current}), the minimal current not only grows,
but also is shifted from $\epsilon_2=0$ to the value $\epsilon_2 =
\lambda^2/\hbar\omega_0$.  This shift can be
obtained by a polaron transformation, as we will detail in
Sec.~\ref{sec:limits}.  This motivates us to henceforth plot the
current as a function of the renormalized detuning
$\epsilon=\epsilon_2-\lambda^2/\hbar\omega_0$. 

\begin{figure}[tb]
\begin{center}
\includegraphics{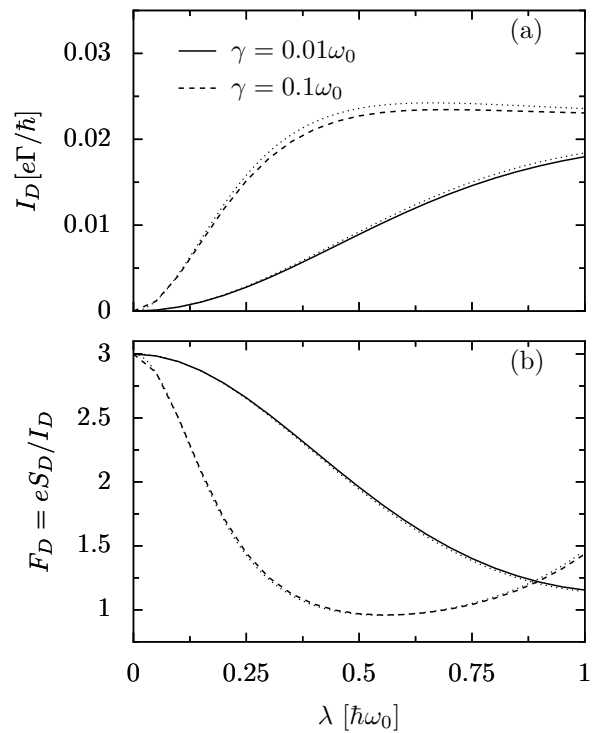}
\end{center}
\caption {\label{Fig.dark-lambda}
Current and Fano factor as a function of the electron-phonon coupling
for the dark state, i.e., for the detuning $\epsilon=\epsilon_2 -
\lambda^2/\hbar\omega_0=0$, at zero temperature and two
values of the dissipation strength $\gamma$.  All other parameters are
as in Fig.~\ref{Fig.current}a.  The dotted lines mark the results
obtained with the reduced master equation \eqref{eq:effdeco}.}
\end{figure}%
\begin{figure}[tb]
\begin{center}
\includegraphics{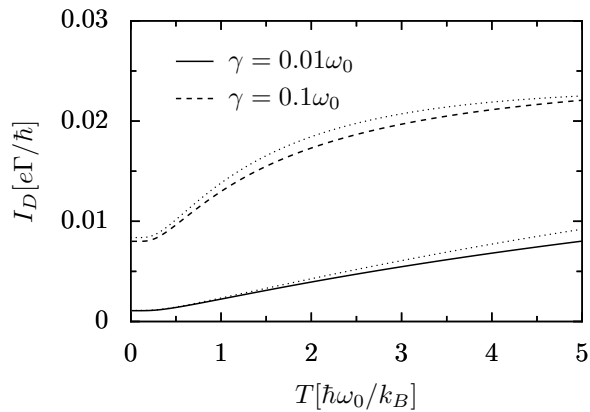}
\end{center}
\caption {\label{Fig.dark-temp}
Current as a function of the temperature for electron-phonon coupling
$\lambda=0.15\hbar\omega_0$ and detuning $\epsilon
=\epsilon_2-\lambda^2/\hbar\omega_0=0$, corresponding to the dark
state. The dotted lines are obtained with the reduced master
equation~\eqref{eq:effdeco} for the dot electrons.  All other
parameters are as in Fig.~\ref{Fig.current}.}
\end{figure}
Figures~\ref{Fig.dark-lambda}a and \ref{Fig.dark-temp} show the
current as a function of the electron-phonon coupling and the
temperature, respectively, for a detuning $\epsilon=0$ which
corresponds to the dark state.  Both plots confirm that the current
blockade is resolved with increasing electron-phonon coupling and
temperature, underlining the growing importance of decoherence.  The
current saturates at the value $I_D \approx 0.02e\Gamma/\hbar$, as a
function of the electron-phonon coupling $\lambda$; see
Fig.~\ref{Fig.dark-temp}a.  A similar behavior has been found for an
interferometer that consists of two quantum dots.\cite{Marquardt2003a}
Figures~\ref{Fig.dark-lambda}b depicts the associated current noise in
terms of the Fano factor.  Starting at the super-Poissonian value
$F=3$, the Fano factor reduces towards $F\approx 1$, indicating a
transition from avalanche-like transport to a Poisson process.

\subsection{Dot 2 far from resonance}

When dot~2 is strongly detuned, i.e., for $|\epsilon_2| \gg \tau$,
tunneling from and to this dot becomes off-resonant.  Then the direct
path from dot~1 to dot~3 is much more likely than the detour via
dot~2.  Then without the oscillator, we expect interference effects to
play a minor role.  Nevertheless, electrons may be trapped in dot~2
such that the current flow is interrupted until the trapped electron
tunnels off-resonantly to dot~3 and transport is restored.
Consequently, the electron transport becomes
bunched.\cite{Dominguez2010a} The current plotted in
Fig.~\ref{Fig.current} demonstrates that this scenario needs to be
refined when the electron on dot~2 couples to a vibrational mode,
because then temporal electron trapping can be caused also by
emission and absorption of phonons.  This leads to dips and peaks in
the current whenever $\epsilon_2$ is detuned by roughly an integer multiple of
$\hbar\omega_0$.  For finite temperature and negative detuning
(Fig.~\ref{Fig.current}b for $\epsilon_2<0$), the dips are caused by
the predominating phonon emission, while those for positive detuning
are due to the a more frequent absorption.  The different size of the
peaks and dips for positive and negative values of $\epsilon$
(Fig.~\ref{Fig.current}b) stems from spontaneous processes which
render emission more likely than absorption.  In the zero temperature
limit (Fig.~\ref{Fig.current}a), phonon absorption no longer occurs
and consequently, the dips at positive detuning vanish.  Then small
peaks emerge, which correspond to the relaxation of electrons that
temporally populate in dot~2.

\section{Elimination of the dissipative phonon}
\label{sec:limits}

In order to obtain a reduced master equation for the triple quantum
dot, we eliminate the phonon via a polaron transformation under a
weak-coupling assumption.\cite{Brandes1999a, Brandes2003a} This
converts the electron-phonon coupling into a renormalized inter-dot
tunneling and additional dissipative terms.  In order to keep
decoherence effects stemming from the phonon-bath coupling, we have to
apply this transformation also to those terms of the master equation
\eqref{eq:TrME} that describe phonon dissipation.

\subsection{Polaron transformation}

We start with the unitary transformation\cite{Mahan1990a,
Brandes1999a} $O \to \bar O = SOS^\dagger$ of the master
equation~\eqref{eq:TrME}, where
\begin{align}
\label{polaron}
S=\exp\left[\frac{\lambda}{\hbar\omega_0}n_2(a^\dagger-a)\right].
\end{align}
This corresponds to the replacements
\begin{align}
\label{polaron:a}
a \to{} & a - \frac{\lambda}{\hbar\omega_0}n_2 ,
\\
\label{polaron:c2}
c_2 \to{} & c_2 X^\dagger
\end{align}
with the phonon displacement operator
\begin{equation}
\label{displacement}
X = \exp\Big[\frac{\lambda}{\hbar\omega_0}(a^\dagger-a) \Big] .
\end{equation}
Notice that all lead and bath operators remain unchanged.  The
Hamiltonian $H_0$ of the dot electrons and the phonon then reads
\begin{equation}
\bar H_{0}
= \epsilon \widehat n_2
  +\tau(c_{1}^\dagger c_3+c_{2}^\dagger c_3X
            +c_{1}^\dagger c_2X^{\dagger} + \text{h.c.})
  +\hbar \omega_0 a^{\dagger}a,
\label{eq:imint}
\end{equation}
where $\epsilon=\epsilon_2-\lambda^2/\hbar\omega_0$ denotes the
effective detuning.

The form \eqref{eq:imint} of the system Hamiltonian allows us to
eliminate the phonon within second-order perturbation theory in the
interdot tunneling.  Then we obtain a master equation for the electron
operators which still depends on electron-oscillator correlations.
Next, the phonon is traced out under the assumption that the polaron
transformation captures most of these correlations, such that the
density operators in the polaron picture factorizes, $\rho = \rho_{e}
\otimes\rho_\text{ph}$.  A similar route has been already taken in
Refs.~\onlinecite{Brandes1999a, Brandes2003a}; it is equivalent to the
non-interacting blip approximation common in quantum
dissipation.\cite{Dekker1987a, Dekker1987b, Morillo1993a}  Here we
only discuss the resulting master equation, while details of the
derivation are provided in Appendix~\ref{app:effective}.

\begin{figure}[tb]
\begin{center}
\includegraphics{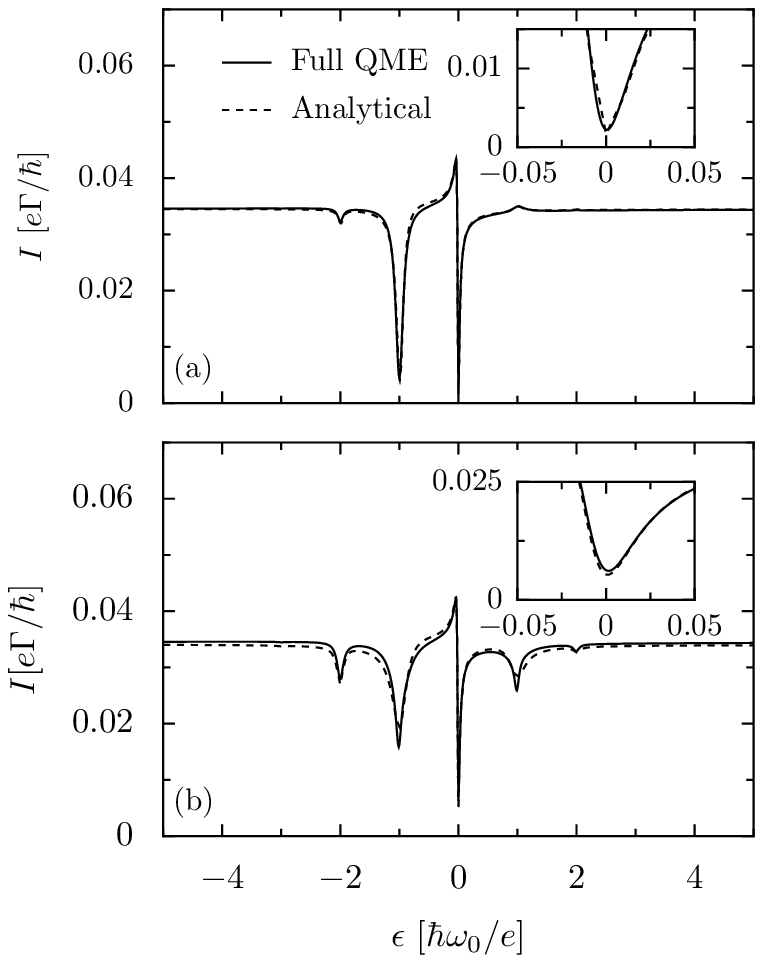}
\end{center}
\caption {\label{Fig.teo} Comparison of the results with the full
quantum master equation \eqref{eq:TrME} and those of the
effective master equation \eqref{eq:effdeco} for
$\lambda=0.1\hbar\omega_0$, $\gamma=0.05\omega_0$,
$\tau=0.01\hbar\omega_0$ and $\Gamma=0.1\omega_0$.  The temperature
is(a) $T=0$ and (b) $T=1.5\hbar\omega_0/k_B$.}
\end{figure}

The resulting quantum master equation contains the effective dot
Hamiltonian
\begin{equation}
\label{Hdot,eff}
H_\text{TQD,eff} = \epsilon n_2
+ \tau (c_1^\dagger c_3 +\text{h.c.})
+ \bar\tau(c_2^\dagger c_3 + c_1^\dagger c_2 +\text{h.c.}) ,
\end{equation}
where the electron tunneling between dot~2 and the two other quantum
dots is renormalized according to
\begin{equation}
\tau \to \bar\tau
= \tau \langle X\rangle=\tau \exp\Big\{ -\Big|\frac{\lambda}{\omega_0}\Big|^2
  \coth\Big(\frac{\hbar\omega_0}{2k_BT}\Big) \Big\}.
\end{equation}
Besides this renormalization, two additional Liouvillians emerge.  The
first one describes decoherence of the dark state, leading to a
small residual current.  It is directly obtained by the replacement
\eqref{polaron:a} in the last two terms of the master equation
\eqref{eq:TrME} and reads
\begin{align}
\mathcal{L}_\text{dec}\rho_e
= \frac{\gamma}{2}(1 + 2\bar n)\Big(\frac{\lambda}{\hbar\omega_0}\Big)^2
  (2 n_2\rho_e  n_2- n_2\rho_e-\rho_e n_2),
\label{eq:decohph}
\end{align}
where we have used the operator relation $n_2^2=n_2$.
We will further analyze the corresponding decoherence mechanism in
Sec.~\ref{sec:dec-mechanism}.  The second Liouvillian stems from the
double commutator in the Bloch-Redfield master equation
\eqref{eq:secorder} and describes incoherent tunneling between the
quantum dots,
\begin{equation}
\begin{split}
\mathcal{L}_\text{ict} \rho_e
=& -\Big(\frac{\tau}{\hbar}\Big)^2 \left\{ \Big(C_{-\epsilon}
       \left(n_1+n_3\right) + 2 n_2 C_{\epsilon} 
    \Big) \rho_e +\text{h.c.} \right\}
\\ &
   +2\Big(\frac{\tau}{\hbar}\Big)^2 \big\{
        C_{-\epsilon}' c_2^\dagger c_3\rho_e c_3^\dagger c_2
       +C_{-\epsilon}' c_2^\dagger c_1\rho_e c_1^\dagger c_2
\\ &
       +C_{\epsilon}' c_1^\dagger c_2\rho_e c_2^\dagger c_1
       +C_{\epsilon}' c_3^\dagger c_2\rho_e c_2^\dagger c_3 \Big\},
\end{split}
\label{eq:liouvcorr}
\end{equation}
where $C_{\epsilon} \equiv C_\epsilon' +iC_\epsilon''$ denotes the
phonon correlation function in Laplace space, derived in
Appendix~\ref{app:corrfunct}.
%
This incoherent inter-dot tunneling is responsible for the current
dips and peaks at the resonances $\epsilon=n\hbar\omega_0$ observed in
Figs.~\ref{Fig.current} and \ref{Fig.teo}.  It occurs with the rates
$2(\tau/\hbar)^2 C_{\epsilon}'$ (between dots 1 and 3) and
$2(\tau/\hbar)^2 C_{-\epsilon}'$ (between dot 2 and dots 1,3),
respectively, which is in accordance with
$P(E)$-theory.\cite{Ingold1992a, Brandes2005a}

In summary, the effective master equation for the triple quantum dot
under the influence of a dissipative phonon and with the coupling to
the leads reads
\begin{equation}
\begin{split}
\dot\rho_e = &
-\frac{i}{\hbar}[H_\text{TQD,eff},\rho_e]+
\mathcal{L}_\text{dec}\rho_e+\mathcal{L}_\text{ict}\rho_e
\\ &
   -\frac{\Gamma_L}{\hbar} (2 c_1\rho_e c_1^{\dagger}-c^{\dagger}_1
    c_1\rho_e-\rho_e c^{\dagger}_1c_1)
\\ &
-\frac{\Gamma_R}{\hbar} (2 c_3\rho_e c_3^{\dagger}-c^{\dagger}_3
   c_3\rho_e-\rho_e c^{\dagger}_3c_3) .
\end{split}
\label{eq:effdeco}
\end{equation}

Numerical calculations provide evidence that $\mathcal{L}_\text{ict}$
is not relevant for the bahavior of the dark state (see
Fig.~\ref{Fig.dark-lambda}).  Thus, close to
$\epsilon=0$, we can neglect $\mathcal{L}_\text{ict}$ in the master
equation \eqref{eq:effdeco}, and then we obtain to lowest order in
$\tau$ the stationary current
\begin{widetext}
\begin{align}
\label{I:polaron}
I_D\approx\frac{4\Gamma\left(4g_1(\tau^2-\bar\tau^2\right)^2+g_1g_2\tau^2\Gamma_D+g_2\bar\tau^2\Gamma\Gamma_D}{\Gamma(2\Gamma+3\Gamma_D)(4g_1\bar\tau^2+g_2\Gamma\Gamma_D)+4\tau^2\left(2\Gamma^3+7\Gamma^2\Gamma_D+12\Gamma\Gamma_D^2+8\Gamma_D^3\right)},
\end{align}
\end{widetext}
with $g_1=\Gamma+2\Gamma_D$, $g_2=\Gamma+\Gamma_D$, and 
the effective dissipation rate $\Gamma_D= (\frac{1}{2} +\bar
n)\gamma(\lambda/\hbar\omega_0)^2$.  The validity of this result close
to the dark state is investigated with
Figs.~\ref{Fig.dark-lambda} and \ref{Fig.dark-temp}.  The agreement is
rather good for any coupling constant $\lambda$ and temperature.
The according result for the Fano factor
also fits well (see Fig.~\ref{Fig.dark-lambda}b).  A comparison in a broad
range of detunings, shown in Fig.~\ref{Fig.teo}, demonstrates that
the approximation is globally valid.

\subsection{Decoherence mechanism}
\label{sec:dec-mechanism}

A physical picture of the electron decoherence can be developed by
considering the influence of the phonon on the dark
state~\eqref{eq:darkwf}.  This reasoning will also yield the
associated decoherence rate of the effective 
Liouvillian~\eqref{eq:decohph}.

Let us assume that the electron resides in the dark state
$|\Psi_{\text{dark}}\rangle \propto |1\rangle-|2\rangle$.  Its time
evolution under the influence of the phonon is determined by the
interaction-picture Hamiltonian
\begin{equation}
\label{eq:intpic}
H_I(t)
=\lambda n_2(t) (a^\dagger e^{i\omega_0 t}+a e^{-i\omega_0 t}) .
\end{equation}
Since the electron dynamics is much slower than the oscillator, the
number operator $n_2$ is essentially time-independent.  Then
the time ordering in the corresponding time-evolution operator
\begin{equation}
U(t)=T_{\leftarrow}\exp\Big[-\frac{i}{\hbar} \int_0^t ds \, H_I(s)\Big]
\label{U.intpic}
\end{equation}
can be evaluated by employing the commutation
relation\cite{Breuer2003a}
\begin{equation}
[H_I(t),H_I(t')]=2i\lambda^2 n_2 \sin[\omega_0 (t-t')]
\end{equation}
from which we obtain the propagator
\begin{equation}
U(t)=\exp\left[-\frac{1}{2}\int_0^t ds \,ds'
     [H_I(s),H_I(s')]\theta(s-s') \right] V(t) .
\label{eq:ev}
\end{equation}
The operator
$V(t)= \exp\{n_2 [a^{\dagger}\alpha(t)-a \alpha(t)^*]\}$
describes an oscillator displacement by
\begin{equation}
\alpha(t)=\frac{\lambda}{\hbar \omega_0} (1-e^{i\omega_0 t}) ,
\end{equation}
while the integral of the commutator in Eq.~\eqref{U.intpic} is a mere
phasefactor which is not relevant for the subsequent discussion and
will be ignored.  Thus, the dark state evolves according to
\begin{equation}
U(t)|\Psi_\text{dark}\rangle
= \frac{1}{\sqrt{2}}\big(|1\rangle|0\rangle_\text{ph}
      -|2\rangle|\alpha(t) \rangle_\text{ph} \big) ,
\label{eq:transitory}
\end{equation}
which means that the oscillator turns into a cat state, i.e., a
superposition of two coherent states.  The coherence of such a state
is known to decay with the rate \cite{Caldeira1985a, Walls1985a}
$\Gamma_D(t) = (\gamma/2)(1+2\bar n)|\alpha(t)|^2$.  For weak
oscillator damping, $\gamma\ll\omega_0$, we can replace the rate by
its time-average \begin{equation} \Gamma_D = \frac{\gamma}{2} (1+2\bar
n) \bigg|\frac{\lambda}{\hbar\omega_0}\bigg|^2 .  \label{eq:decrate}
\end{equation} Notice that we do not trace out the electrons, but
consider the coherence of the electron-phonon compound.

Since each of the two involved phonon states is linked to a particular
electron state, we can attribute this decoherence process also to the
electrons.  Then we can conclude that the electron coherence also
decays with the rate~\eqref{eq:decrate}, which complies with the
actual rate in the effective Liouvillian Eq.~\eqref{eq:decohph}.
Thus, the phonon elimination described above is such that the
decoherence of an oscillator cat state directly turns into decoherence
of the dark state.

For larger inter-dot tunneling, $\tau\gtrsim\hbar\omega_0$, the
interaction-picture operator $n_2(t)$ can no longer be considered
time-independent, such that our reasoning has to be modified.
Moreover, if we would use a model in which also dot~1 couples to the
phonon, the dark electron state and the phonon state would factorize
and be $\propto (|1\rangle-|2\rangle) |\alpha(t)\rangle$.  Then no
phonon-induced decoherence would take place and, consequently, the
dark state would continue to block the electron transport.

\section{Conclusions}

We have investigated decoherence effects in a triple quantum dot
interferometer the stemming from the coupling to a single dissipative
bosonic mode.  In our model, the dots are arranged in a symmetric ring
configuration in which two dots couple to source and drain, while the
third dot interacts with a dissipative harmonic oscillator.  In the
absence of the oscillator, a strong detuning of the third dot leads to
electron trapping and bunching.  When all dots are close to resonance,
by contrast, interference effects dominate.  In particular, ideal
destructive interference may occur, such that the current vanishes
completely, even when all electronic energy levels lie within the
voltage window.

It turned out that the oscillator entails two effects: First, the
current minimum is found at a shifted detuning and, second,
destructive interference is no longer perfect, such that always a
finite current emerges.  This suspension of destructive interference
is also visible in the current noise measured in terms of the Fano
factor.  When the residual current is very small, i.e., for small
decoherence, the associated shot noise is enhanced, while transport
becomes almost Poissonian with stronger decoherence.

A qualitative understanding of these effects has been achieved by an
analytical approximation after a polaron transformation leading to a
reduced master equation for only the dot electrons.  Within a standard
treatment similar to the non-interacting blip approximation, we have
obtained an effective master equation for the electron transport.
Then it became possible to analytically obtain the current from the
resulting master equation also close to destructive interference.  The
results agree well with the full numerical results, provided that the
oscillator frequency is sufficiently large and the intra-dot tunneling
is small.  In turn, we can conclude that our reduced master equation
faithfully describes transport effects entailed by a dissipative mode.
Moreover, this picture provide evidence that the decoherence of an
oscillator cat state directly turns into decoherence of the dark
state.

In summary, our results underline the impact of already one phonon
mode on quantum dot interferometers.  With our reduced master equation
for the quantum dot electrons, we have put forward a method for
describing such systems efficiently after eliminating the oscillator.
Such a method is in particular welcome when the oscillator is only
weakly damped, since then an explicit treatment requires taking quite
a few oscillator states into account.

\acknowledgments
We like to thank P. C. E. Stamp for enlightening discussions.  This
work has been supported by the Spanish Ministry of Science and
Innovation through project MAT2008-02626, via a FPI grant (F.D.),
and by the European project ITN under Grant No.\ 234970 EU.

\appendix
\section{Effective master equation}
\label{app:effective}

In this appendix we provide some details of the derivation of the
effective master equation~\eqref{eq:effdeco} starting with the
polaron-transformed electron-phonon Hamiltonian~\eqref{eq:imint}.  We
treat all terms that couple dot~2 to the phonon within second order
perturbation theory, which means that we separate the electron-phonon
Hamiltonian as $H_\text{TQD,eff} + H_Y$, where $H_\text{TQD,eff}$ is defined 
in Eq.~(\ref{Hdot,eff}) and 
\begin{equation}
H_Y(t) = \tau ( c_2^\dagger c_3 Y_t
   +c_1^\dagger c_2 Y_t^{\dagger}+\text{h.c.} ).
\label{eq:ht}
\end{equation}
The latter Hamiltonian will be treated within Bloch-Redfield approximation.
The phonon part of the interaction, $Y = X-\langle X\rangle_\text{eq}$,
 has been defined such that $\langle H_Y\rangle_\text{eq}$ vanishes.
Then within the usual Born approximation,\cite{Blum1996a} we obtain in the
interaction picture the master equation
\begin{equation}
\begin{split}
\frac{d}{dt} \tilde\rho(t)
=& -\frac{1}{\hbar^2}\int_0^{t}ds \big[\tilde H_Y(t),
   \big[\tilde H_Y(s),\tilde \rho(s)\big]\big] ,
\end{split}
\label{eq:secorder}
\end{equation}
where the contribution of first order in the perturbation $H_Y$
vanishes owing to $\langle H_Y\rangle_{\text{eq}}=0$.  A simplification of the
master equation \eqref{eq:secorder} comes from the fact that its
right-hand side is already of second order in the inter-dot tunneling
$\tau$, while higher orders are neglected.  It is therefore sufficient
to depricate in the interaction-picture representation of $H_Y$ the
tunneling terms in $H_\text{TQD,eff}$, such that the corresponding unperturbed
propagator reads $U_0' = \exp(-i\epsilon n_2 t/\hbar)$.

If the electron-phonon interaction is much smaller than the phonon
energy, $\lambda\ll\hbar\omega_0$, the correlation between these two
subsystems is by and large captured by the polaron
transformation.  Thus, in the polaron picture, we can evaluate the
master equation under the factorization assumption $\tilde\rho(t)
\approx \rho^0_\text{ph} \text{Tr}_{\text{ph}}\tilde \rho(t')$.
This corresponds to a non-interacting blip approximation
\cite{Dekker1987a, Dekker1987b, Morillo1993a} for a dissipative
quantum system and has been used also to eliminate a single
dissipative phonon in the context of both quantum
transport \cite{Brandes2003a, Brandes1999a} and quantum
dissipation.\cite{Nesi2007a}  

Within Born approximation, it is consistent to replace in the master
equation \eqref{eq:secorder} the time arguments of the density matrix
by the final time $t$.  When finally tracing out the phonon, we
obtain expectation values of the type
\begin{align}
&c^\dagger_{i}c_j(t)\rho(t) c^\dagger_{i}c_j(s)\langle X^\dagger_t X_s \rangle,\label{eq:term2}\\
&c^\dagger_{i}c_j(t)\rho(t) c^\dagger_{i}c_j(s)\langle X_t X_s \rangle,\label{eq:neg1}\\
&c^\dagger_{i}c_2(t)\rho(t) c^\dagger_{2}c_j(s)\langle X^\dagger_t X_s \rangle.\label{eq:neg2}
\end{align}
Terms of the type \eqref{eq:term2}
give rise to the additional Liouvillian \eqref{eq:liouvcorr}.  The two
following terms are negligible for different reasons.  The term
\eqref{eq:neg1} depends on the time $t+s$ and, thus, is rapidly oscillating.
Therefore it can be neglected within a rotating-wave approximation.
Finally, terms of the type~\eqref{eq:neg2} come in pairs with opposite
time-ordering and opposite sign.  Therefore their net contribution is
proportional to a commutator and, thus, is of the order $\tau$, i.e.,
one order beyond what is considered in the master equation
\eqref{eq:secorder}.

\section{Correlation function}
\label{app:corrfunct}

The effective Liouvillian derived in Appendix~\ref{app:effective} contains
averages over one and two phonon displacement operators.  We calculate
them using the quantum regression theorem which is valid within Markov
approximation. \cite{Breuer2003a} The renormalization of the coherent
tunneling stems from averages of the type
\begin{align}
c_i^\dagger c_j \langle X_t \rangle&=c_i^\dagger c_j \text{Tr}_\text{ph}\left\{ X \rho_\text{ph}(t) \right\}\nonumber\\
&=c_i^\dagger c_j\text{Tr}_\text{ph}\left\{ X \rho_\text{ph,eq} \right\}\nonumber\\
&= c_i^\dagger c_j \exp\Big\{ -\Big|\frac{\lambda}{\omega_0}\Big|^2
  \coth\Big(\frac{\hbar\omega_0}{2k_BT}\Big) \Big\} ,
\end{align}
with the equilibrium phonon density matrix
\begin{equation}
\rho_\text{ph,eq} = \frac{1}{Z} \exp(-\hbar\omega_0 a^{\dagger}a/k_BT),
\end{equation}
and the partition sum $Z=[1-\exp(-\hbar\omega_0/k_BT)]^{-1}$.

Using once more the quantum regression theorem, we write
the correlation function as
\begin{equation}
C(t) = \langle X^{\dagger}(0) X(t)\rangle_\text{eq}
= \tr\{ X^{\dagger}(0) X_H(t) \rho_\text{ph,eq}\} ,
\label{C(t)}
\end{equation}
i.e., with a Heisenberg operator that fulfills the
equation of motion $\dot a_H =
-(i\omega_0+\gamma/2)a_H$.  From its solution
\begin{equation}
a_H= a e^{-(i\omega_0+\gamma/2)t}
\end{equation}
follows the displacement operator in the interaction picture,
\begin{equation}
X_t \equiv X(t) = \exp \left[
\frac{\lambda}{\omega_0}\left(a^{\dagger}e^{(i\omega_0-\gamma/2)t}
- \text{h.c.} \right) \right] .
\end{equation}
Inserting this operator and $\rho_\text{ph,eq}$ into the correlation
function~\eqref{C(t)} yields
\begin{equation}
\label{eq:correlatorT}
\begin{split}
C(t)
= &\exp\bigg[ \Big|\frac{\lambda}{\omega_0}\Big|^2
   \bigg\{ie^{-\gamma/2 t}\sin(\omega_0 t)
         + \coth\Big(\frac{\hbar\omega_0}{2k_BT}\Big)
\\ &\times \left(1+e^{-\gamma t}-2e^{-\gamma/2 t}
     \cos(\omega_0 t)\right)\bigg\}\bigg] .
\end{split}
\end{equation}
For computing the coefficients of the master equation, we need this
correlation function in Laplace space, evaluated at $z=0$, defined as
\begin{align}
C_\epsilon
= \lim_{z\rightarrow0}\int_0^\infty dt\,
  e^{-(z+i\epsilon)t/\hbar} C(t)
\equiv C_\epsilon' + iC_\epsilon''
.
\end{align}


\end{document}